\documentclass[aps,pre,twocolumn,english,superscriptaddress,showpacs,floatfix]{revtex4}
\usepackage{amsmath}
\usepackage{amssymb}
\usepackage{epsfig,graphics,graphicx}
\usepackage{babel}

\begin{document}

\title{Mesoscopic Model for Diffusion-Influenced Reaction Dynamics}

\author{Kay Tucci}
\email[]{kay@ula.ve} \affiliation{Max-Planck-Institut f\"ur Physik
Komplexer Systeme, N\"othnitzer Strasse 38, 01187 Dresden,
Germany} \affiliation{SUMA-CeSiMo, Universidad de Los Andes,
M\'erida 5101, Venezuela}
\author{Raymond Kapral}
\email[]{rkapral@chem.utoronto.ca}
\affiliation{Max-Planck-Institut f\"ur Physik Komplexer Systeme,
N\"othnitzer Strasse 38, 01187 Dresden, Germany}
\affiliation{Chemical Physics Theory Group, Department of
Chemistry, University of Toronto, Toronto, ON M5S 3H6, Canada}

\date{\today}

\begin{abstract}
A hybrid mesoscopic multi-particle collision model is used to
study diffusion-influenced reaction kinetics. The mesoscopic
particle dynamics conserves mass, momentum and energy so that
hydrodynamic effects are fully taken into account. Reactive and
non-reactive interactions with catalytic solute particles are
described by full molecular dynamics. Results are presented for
large-scale, three-dimensional simulations to study the influence
of diffusion on the rate constants of the $A+C \rightleftharpoons
B+C$ reaction. In the limit of a dilute solution of catalytic C
particles, the simulation results are compared with diffusion
equation approaches for both the irreversible and reversible
reaction cases. Simulation results for systems where the volume
fraction $\phi$ of catalytic spheres is high are also presented,
and collective interactions among reactions on catalytic spheres
that introduce volume fraction dependence in the rate constants
are studied.
\end{abstract}


\maketitle

\section{Introduction}

The dynamics of large complex systems often occurs on disparate
time and space scales. The direct molecular dynamics simulation of
the equations of motion for such systems is difficult because of
this scale separation and the large numbers of molecules such
systems may contain. Consequently, mesoscopic models play an
important role in investigations of the dynamics of these systems.

The use of Langevin and Fokker-Planck equations for Brownian
motion is well known \cite{chand,gardiner} and these models have
been used in much wider contexts; for example, in investigations
of reaction dynamics in the condensed phase \cite{hanggi}. Such
stochastic models are useful when it is impossible or
inappropriate to simulate the full dynamics of the system,
including all solvent degrees of freedom.

Suspensions of colloidal particles are also often treated using
mesoscopic models of various types. While the dynamics of the
colloidal particles may be accurately modelled using Langevin
dynamics, hydrodynamic interactions play an important role in
dense colloidal suspensions. The friction tensors that enter the
Langevin equations depend on the colloidal particle configuration.
To compute the frictional properties of dense suspensions, the
intervening solvent is often approximated by the continuum
equations of hydrodynamics to determine the hydrodynamic
interactions among the colloidal particles.

Other approaches for constructing mesoscopic dynamics of complex
systems include the construction of effective solvent models to be
used in the context of full molecular dynamics simulations.
\cite{steve} Such models allow one to investigate systems of high
complexity that cannot be studied by straightforward molecular
dynamics simulation schemes.

In this article show how diffusion-influenced reactions can be
studied using a multi-particle mesoscopic dynamics.
\cite{stostream,meso1} In this dynamical scheme, particle
positions and velocities are continuous variables and the dynamics
consists of free streaming and multi-particle collisions.
Multi-particle collisions are carried out by partitioning the
system into cells and performing a specific type of random
rotation of the particle velocities in each cell that conserves
mass, momentum and energy. The hydrodynamic equations are obtained
on long distance and time scales \cite{meso1} and the model
permits efficient simulation of hydrodynamic flows
\cite{meso1,kroll}. Since the dynamics is carried out at the
particle level, it is straightforward to construct hybrid schemes
where solute molecules that undergo full molecular dynamics are
embedded in the mesoscopic solvent. \cite{meso2} Hydrodynamic
interactions among solute particle are automatically accounted for
in the multi-particle mesoscopic dynamics. \cite{poly} The method
has been generalized to treat phase segregating fluids with
surfactants. \cite{chen}

Diffusion-influenced reaction dynamics is widely used to model
processes like enzymatic turnover or collision-induced
isomerization in complex systems. Smoluchowski constructed a
continuum theory for such reactions based on a solution of the
diffusion equation. \cite{smol} In this article we focus on the
reversible $A+C \rightleftharpoons B+C$ reaction where a
considerable body of research has concerned the development of
refined theoretical models
\cite{pagitsas,lee,agmon1,naumann,gopich,yang,gopich2}. Simulation
schemes \cite{kim,oh,popov,agmon2} for three-dimensional diffusive
reaction dynamics have been constructed. Diffusion-influenced
reactions taking place in a dense field of catalytic particles are
strongly affected by perturbations of the diffusion field arising
from reactions at the different catalytic sites.
\cite{felderhof1,lebenhaft,felderhof2,felderhof3,gopich3,
felderhof4,gopich4} This effect is similar to the hydrodynamic
interactions that enter colloidal suspension dynamics. We show how
collective effects on diffusion-influenced reaction dynamics can
be studied by simulations of a mesoscopic model for these systems.
The mesoscopic multi-particle collision model allows us to
simulate systems with tens of millions of particles for long times
in order to determine power law decays and non-analytic catalytic
particle density effects on the reaction rates.

The outline of the paper is as follows. Section~\ref{sec:multiCol}
sketches the mesoscopic multi-particle collision model and
presents its generalization to multi-component systems. The
evolution equations that encode the multi-particle mesoscopic
dynamics are presented in Sec.~\ref{sec:evoleq}. The computation
of the diffusion coefficient, a necessary ingredient for the
analysis of reaction dynamics, is given in
Sec.~\ref{sec:diffusion}. In Sec.~\ref{sec:reaction} we show how
the model can be generalized to treat chemical reactions. In
particular we study the reaction $A +C \rightleftharpoons B +C$
that occurs upon collision with catalytic C particles. The
simulation algorithms and simulation results for dilute and
concentrated suspensions of catalytic spheres are presented in
Sec.~\ref{sec:sim}. The conclusions of the investigation are
contained in Sec.~\ref{sec:conc}.

\section{Multi-Component Mesoscopic Multi-Particle Dynamics} \label{sec:multiCol}

The mesoscopic dynamics we consider comprises two steps:
multi-particle collisions among the particles and free streaming
between collisions. \cite{meso1} Suppose the system contains $N$
particles with positions and velocities given by $({\bf
X}^{(N)},{\bf V}^{(N)})=({\bf x}_1,\dots, {\bf x}_N,{\bf
v}_1,\dots,{\bf v}_N)$. While the particle positions and
velocities are continuous variables, for the purpose of effecting
collisions, the system is divided into $L$ cells labelled by the
index $\xi$. Collisions occur locally in the cells in the
following way: Rotation operators $\hat{\omega}$, chosen randomly
from a set of rotation operators $\Omega=\{\hat{\omega}_1, \dots,
\hat{\omega}_k\}$ are assigned to each cell $\xi$ of the system.
If a cell $\xi$ contains $n_{\xi}$ particles at time $t$ and the
center of mass velocity in the cell is
$\mathsf{V}_{\xi}(t)=n_{\xi}^{-1}\sum_{i=1}^{n_{\xi}} {\bf
v}_i(t)$, the post-collision values of the velocities of the
particles in the cell, ${\bf v}_i^*$, are computed by rotating the
particle velocities relative to $\mathsf{V}_{\xi}$ and adding
$\mathsf{V}_{\xi}$ to the result,
\begin{equation}
{\bf v}_i^*(t)=\mathsf{V}_{\xi}(t)+\hat{\omega}_{\xi}\Big({\bf
v}_i(t)-\mathsf{V}_{\xi}(t)\Big)\;.
\end{equation}
After the collision events in each cell, the particles free stream
to their new positions at time $t+\tau$,
\begin{equation}
{\bf x}_i(t+\tau)={\bf x}_i(t)+{\bf v}_i^*(t) \tau \;.
\end{equation}
This simple dynamics has been shown to conserve mass, momentum and
energy. The exact hydrodynamic equations are obtained on
macroscopic scales, and the system relaxes to an equilibrium
Boltzmann distribution of velocities. \cite{meso1} Consequently,
the dynamics, although highly idealized, has correct behavior on
macroscopic scales which are long compared to the effective
collision times in the model. Since the dynamics is described at
the particle level it is a simple matter to couple this mesoscopic
dynamics to full molecular dynamics of solute species embedded in
it. \cite{meso2,mesofin} The model is similar in spirit to Direct
Simulation Monte Carlo \cite{bird} but with a different
discrete-time collision dynamics that simplifies the simulations
and makes them more efficient.

The mesoscopic dynamics for a multi-component system can be
carried out in a similar way by generalizing the multi-particle
collision rule. Suppose the $N$-particle system comprises
different species $\alpha = A,B,\dots$ with masses $m_\alpha$. In
this case it is useful to introduce an operator $\Theta^\alpha_i$
that characterizes the species $\alpha$ of a given particle $i$.
These operators have the following properties:
\begin{equation}
\Theta^\alpha_i \Theta^{\alpha^\prime}_i= \delta_{\alpha
\alpha^\prime}\; ;
\end{equation}
i.e., particle $i$ cannot be of different species at the same
time; also,
\begin{equation}
\sum_\alpha \Theta^\alpha_i = 1\; ,
\end{equation}
so that particle $i$ has to have some species type. The number of
particles of species $\alpha$ is given by
\begin{equation}
N_\alpha = \sum_{i=1}^N \Theta^\alpha_i\; .
\end{equation}

There are many ways in which the multi-particle collision rule can
be generalized for systems with several species and we consider
one version that is consistent with the requirements that mass,
momentum and energy be conserved. Let $\mathsf{V}^{(\alpha)}_\xi$
be the center of mass velocity of particles of species $\alpha$
that are in the cell $\xi$ at time $t$,
\begin{equation}
\mathsf{V}^{(\alpha)}_\xi(t) = \frac{1}{n^{(\alpha)}_\xi(t)}
     \sum_{i|\mathbf{x}\in\cal V} \Theta^\alpha_i {\mathbf v}_i(t)\; ,
\end{equation}
where $n^{(\alpha)}_\xi$ is the number of particles of the species
$\alpha$ in cell $\xi$ with volume $\cal V$ at time $t$. The
center of mass velocity of all $n_\xi(t)=\sum\limits_\alpha
n_\xi^{(\alpha)}(t)$ particles in the cell $\xi$ at time t is
given by
\begin{equation}
\mathsf{V}_\xi(t) =  \frac {\sum\limits_\alpha n^{(\alpha)}_\xi
m_\alpha  \mathsf{V}^{(\alpha)}_\xi(t)}
    {\sum\limits_\alpha n^{(\alpha)}_\xi m_\alpha} \;.
\end{equation}

In the model we adopt, two different types of multi-particle
collisions occur. The first is a collision that involves particles
of all species. To perform this collision, we use a rotation
operator $\hat{\omega}$ which is applied to every particle in a
cell as for single component system. The second type of
multi-particle collision involves only particles of the same
species. The rotation operator $\hat{\omega}^{\alpha}$ effects
this collision and is applied to each particle of species $\alpha$
in the cell. Not only does it change from cell to cell and with
time like $\hat{\omega}$, but it also changes from species to
species.

The multi-particle collision process can be divided into these two
independent steps. For the set of particles that are in the cell
$\xi$, first we perform the all-species collision as
\begin{equation}\label{eq:collisionAll}
\mathbf{v}_i^{\prime\prime} = \mathsf{V}_\xi
       + \hat{\omega}_\xi({\mathbf v}_i - {\mathsf V}_\xi) \; ,
\end{equation}
where ${\mathbf v}_i$ the pre-collision velocity of the particle
$i$ and $\mathbf{v}_i^{\prime\prime}$ is the velocity after this
step. Second, we apply the one-species rotation operator
\begin{equation}\label{eq:collisionAlpha}
\mathbf{v}_i^* = \sum_\alpha \Theta^\alpha_i \left(
       \mathsf{V}_\xi^{\prime\prime (\alpha)} +
       \hat{\omega}^{\alpha}_\xi({\bf v}_i^{\prime\prime}
       - \mathsf{V}_\xi^{\prime\prime (\alpha)}) \right) \; ,
\end{equation}
where $\mathsf{V}_\xi^{\prime\prime (\alpha)}$ is the center of
mass velocity of particles of species $\alpha$ after the
all-species collision step. Note that $\hat{\omega}_\xi$ is
applied to all particles in the cell, but the
$\hat{\omega}^{\alpha}_\xi$ are applied only on particles of
species $\alpha$.

From Eqs.~(\ref{eq:collisionAll}) and (\ref{eq:collisionAlpha})
the post-collision velocity of a particle may be expressed as
\begin{equation}\label{eq-collision}
\mathbf{v}_i^{*} = \mathsf{V}_\xi
 + \hat{\omega}_\xi(\Theta^{\alpha}_i
   \mathsf{V}_\xi^{(\alpha)} - \mathsf{V}_\xi)
 + \sum_\alpha \Theta^\alpha_i \left(
   \hat{\omega}^{\alpha}_\xi \hat{\omega}_\xi
       ( {\mathbf v}_i - {\mathsf V}_\xi^{(\alpha)}) \right)\; .
\end{equation}

\section{Evolution Equations} \label{sec:evoleq}

The dynamics described above can be encoded in an evolution
equation for the phase space probability density,
\begin{eqnarray}\label{eq:phaseProbDen}
&&{\mathsf P}(\mathbf{V}^{(N)},\mathbf{X}^{(N)} + \mathbf{V}^{(N)}
\tau, t + \tau) \nonumber \\
&& \qquad \qquad \qquad = e^{\mathsf{L}_0\tau} {\mathsf
P}(\mathbf{V}^{(N)},\mathbf{X}^{(N)},t + \tau)\nonumber \\
&&\qquad \qquad \qquad= \hat{\cal C}{\mathsf
P}(\mathbf{V}^{(N)},\mathbf{X}^{(N)}, t) \;,
\end{eqnarray}
where the free streaming Liouville operator is,
\begin{equation}
  \mathsf{L}_0 = \sum_\alpha \sum_{i=1}^N
  \Theta^\alpha_i \left({\mathbf v}_i \cdot {\mathbf \nabla}_i\right) \; ,
  \label{eq:freest}
\end{equation}
and $N = \sum\limits_\alpha N_\alpha$ is the total number of
particles in the system. If we choose the rotation operators
$\hat{\omega}$ and $\hat{\omega}^{\alpha}$ randomly from the set
$\Omega$, the collision operator may be written as,
\begin{eqnarray}
&&\hat{\cal C}{\mathsf P}(\mathbf{V}^{(N)},\mathbf{X}^{(N)}, t) =
  \frac{1}{\|\Omega\|^L} \sum\limits_{\Omega^L}
  \int\limits d\mathbf{V}^{\prime(N)}
  \nonumber \\
&&  \times P(\mathbf{V}^{\prime(N)}, \mathbf{X}^{(N)},
t)\prod_\alpha \prod\limits_{i=1}^N \Theta^\alpha_i \ \delta
  \Big( \mathbf{v}_i - \mathsf{V}_\xi^\prime
    \nonumber \\
&& \qquad
   - \hat{\omega}_\xi( \mathsf{V}_\xi^{\prime (\alpha)}-\mathsf{V}_\xi^\prime)
    - \hat{\omega}^\alpha_\xi \hat{\omega}_\xi
    ( \mathbf{v}_i^{\prime} - {\mathsf V}_\xi^{\prime(\alpha)})\Big)
    \;,
\end{eqnarray}
where $L$ is the number of cells.

We may write the evolution equation in continuous time by
introducing a delta function collision term which accounts for the
fact that the multi-particle collisions occur at discrete time
intervals. We have
\begin{equation}
\frac{\partial}{\partial t} P(\mathbf{V}^{(N)}, \mathbf{X}^{(N)},
t) =
  \left(-{\mathsf L}_0 + {\cal C}\right)
  P(\mathbf{V}^{(N)}, \mathbf{X}^{(N)}, t) \; ,
  \label{eq:ctp}
\end{equation}
where the collision operator ${\cal C}$ acts on the velocities of
the particles at discrete times $m\tau$, and is defined as
\begin{equation}
{\cal C}P(\mathbf{V}^{(N)}, \mathbf{X}^{(N)}, t) =
\sum_{m=0}^\infty \delta(t - m \tau)(\hat{\cal C} - 1)
    P(\mathbf{V}^{(N)}, \mathbf{X}^{(N)}, t) \; .
\label{eq:calC}
\end{equation}
If Eq.~(\ref{eq:ctp}) is integrated over a time interval $m\tau
-\epsilon$ to $(m+1)\tau -\epsilon$ we recover
Eq.~(\ref{eq:phaseProbDen}) corresponding to multi-particle
collision followed by free streaming. Instead, integration over
the jump at $t=(m+1)\tau$ yields an analogous discrete time
equation with free streaming followed by collision.

Assuming that the system is ergodic, then, in view of the
conservation of mass, momentum and energy, the stationary
distribution of the Markov chain in Eq.~(\ref{eq:phaseProbDen}) is
given by the microcanonical ensemble expression,
\begin{eqnarray}
  P_0(\mathbf{V}^{(N)}, \mathbf{X}^{(N)}) &=&
  {\cal N}\delta\left(\frac{1}{2N} \sum_{i=1}^N
  \sum\limits_\alpha \Theta^\alpha_i m^\alpha\|{\mathbf v}_i \|^2
  - \frac {d}{2 \beta} \right)\nonumber \\
  &&  \times
  \delta \left(\sum_{i=1}^N\sum\limits_\alpha \Theta^\alpha_i m^\alpha\
  ({\mathbf v}_i - \bar{\mathbf v})\right)\; ,
\end{eqnarray}
where $\bar{\mathbf v}$ is the mean velocity of the system, $d$ is
the dimension and ${\cal N}$ is a normalization constant. If we
integrate $P_0$ over the phase space of all particles except
particle $i$, we obtain the Maxwell-Boltzmann distribution in the
limit of large $N$.

\begin{figure}[htbp]
\centerline{\mbox{
\epsfig{file=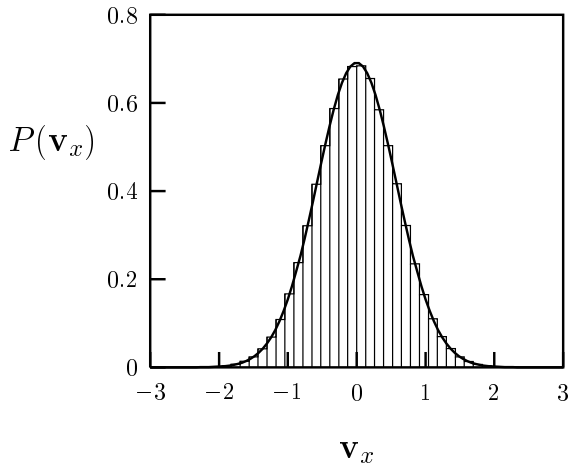,width=0.75\linewidth,clip=,angle=90} }}
\caption{Comparison of the simulated velocity distribution
(histogram) with the Maxwell-Boltzmann distribution function
(solid line) for $k_BT=1/3$.
         \label{fig-velDist}}
\end{figure}

Figure~\ref{fig-velDist} shows the results of a simulation of the
velocity probability distribution for a system with volume $V =
100^3$ cells of unit length and $N=10^7$ particles. The particles
were initially uniformly distributed in the volume $V$ and all
particles had the same speed $|{\mathbf v}| = 1$ but different
random directions. To obtain the results in this figure we assumed
that the species were mechanically identical with mass $m=1$ and
used the multi-particle collision rule in Eq.~(\ref{eq-collision})
with rotations $\hat{\omega}_\xi$ and $\hat{\omega}_\xi^\alpha$
selected from the set $\Omega=\{\pi/2,-\pi/2\}$ about axes whose
directions were chosen uniformly on the surface of a sphere ($\pm
\pi/2$ collision rule). This version of the collision rule for
mechanically identical particles will be used in all calculations
presented in this paper.
The figure compares the histogram of the x-component of the
velocity with the Maxwell-Boltzmann distribution,
\begin{equation}
  P_m(v_x) =
  \left(\frac{m \beta}{2 \pi} \right)^{1/2} e^{-\beta m
       v_x^2/2} \; ,
\end{equation}
where $\beta = (k_BT)^{-1}$, and confirms that this initial
distribution evolves to the Maxwell-Boltzmann distribution under
under the mesoscopic dynamics.

We may also write an evolution equation for any dynamical variable
$a(\mathbf{V}^{(N)},\mathbf{X}^{(N)})$ as,
\begin{equation}
\frac{d}{d t} a(\mathbf{V}^{(N)}, \mathbf{X}^{(N)}, t) =
  \left({\mathsf L}_0 + \mathsf{C} \right)
  a(\mathbf{V}^{(N)}, \mathbf{X}^{(N)}, t) \; ,
  \label{eq:ctdv}
\end{equation}
where $\mathsf{C}$ has the same form as ${\mathcal C}$ in
Eq.~(\ref{eq:calC}) with $\hat{\mathcal C}$ replaced by
$\hat{\mathsf C}$,
\begin{eqnarray}
&&\hat{\mathsf C}a(\mathbf{V}^{(N)},\mathbf{X}^{(N)}, t) =
  \frac{1}{\|\Omega\|^L} \sum\limits_{\Omega^L}
  \int\limits d\mathbf{V}^{\prime(N)}
  \nonumber \\
&&  \times a(\mathbf{V}^{\prime(N)}, \mathbf{X}^{(N)},
t)\prod_\alpha \prod\limits_{i=1}^N \Theta^\alpha_i \ \delta
  \Big( \mathbf{v}_i^\prime - \mathsf{V}_\xi
    \nonumber \\
&& \qquad  - \hat{\omega}_\xi(
\mathsf{V}_\xi^{(\alpha)}-\mathsf{V}_\xi)
    - \hat{\omega}^\alpha_\xi \hat{\omega}_\xi
    ( \mathbf{v}_i - {\mathsf V}_\xi^{(\alpha)})\Big)
    \;.
\end{eqnarray}
This equation is the starting point for the generalization to
reacting systems in Sec.~\ref{sec:reaction}.

\section{Diffusion} \label{sec:diffusion}

A knowledge of the value of the diffusion coefficient is essential
for the analysis of diffusion-influenced reaction kinetics. In
this section we determine the diffusion coefficient as a function
of the density from simulations of the mesoscopic multi-particle
dynamics and derive an approximate analytical expression for its
value.

\begin{figure}[htbp]
\centerline{\mbox{
\epsfig{file=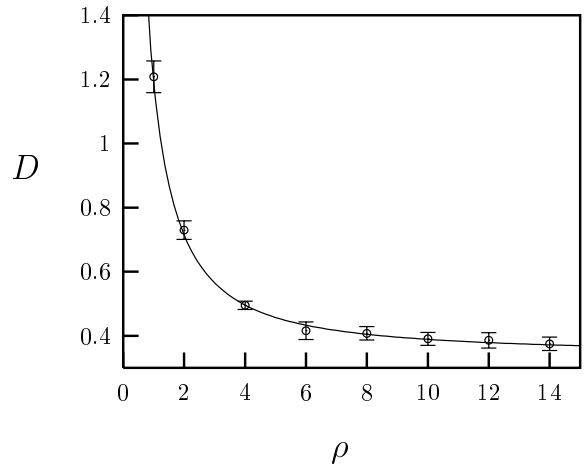,width=0.75\linewidth,clip=,angle=90} }}
\caption{Comparison of the simulated diffusion coefficient
($\odot$) with the Boltzmann value (solid line). The $\pm \pi/2$
collision rule was used to obtain the results. The volume was
$V=100^3$ and the temperature was $k_BT=1/3$.
          \label{fig-DvsRho}}
\end{figure}

The diffusion coefficient is given by the time integral of the
velocity correlation function. For the discrete time dynamics of
the model, the time integral is replaced by its trapezoidal rule
approximation, as shown by a discrete time Green-Kubo analysis.
\cite{meso2,mesofin} Thus, the diffusion coefficient $D$ is given
by
\begin{equation}\label{eq-VAF}
D=\frac{1}{2}\langle v_x v_x \rangle +
  \sum_{\ell=1}^{\infty}\langle v_x v_x(\ell\tau) \rangle \; ,
\end{equation}
where $v_x$ is the x-component of the velocity of a tagged
particle in the system. (We suppress the species index $\alpha$
for the case of mechanically identical particles since all species
have the same diffusion coefficient.) We have computed $D$ using
this expression as well as the formula for $D$ in terms of the
mean square displacement as a function of the mean particle
density per cell $\rho$. The results are shown in
Fig.~\ref{fig-DvsRho}.

An approximate expression for $D$ can be derived by assuming a
single relaxation time approximation. If we suppose the decay is
given by a single relaxation time, we have
\begin{equation}\label{eq-relax}
\frac{\langle v_x v_x(\ell \tau) \rangle}{\langle v_x v_x \rangle}
\approx \left(\frac{\langle v_x v_x(\tau) \rangle}{\langle v_x v_x
\rangle} \right)^\ell \equiv {(r_D)}^\ell \; .
\end{equation}
The diffusion coefficient is then approximately given by
\begin{equation} D\approx-\frac{1}{2}\langle v_x v_x \rangle +
        \langle v_x v_x \rangle \sum_{\ell=0}^{\infty}{r_D}^\ell
 = \frac{\langle v_x v_x \rangle (1+r_D)}{2 (1-r_D)} \; .
 \label{eq-diff}
\end{equation}

The relaxation rate may be computed in the Boltzmann approximation
\cite{meso1},
\begin{eqnarray}
\langle v_{1 x} v_{1 x}(\tau) \rangle &=& \int d{\mathbf v} v_{1x}
  \sum_\omega \sum_{n=1}^\infty \frac{\rho^n}{||\Omega||n!}e^{-\rho}
  \int d{\mathbf v}^{(n)} \nonumber \\
  && \times \delta({\mathbf v} - {\mathbf
  v}_1)  \prod_{i=1}^n \phi({\mathbf v}_i) \sum_{j=1}^n v_{j x}^* \; ,
\end{eqnarray}
where, $v_{1 x}$ is the $x$ component velocity of the single
particle $1$. Since cross correlations between different particles
are not present for self diffusion, we have,
\begin{equation}\label{eq-v1xv1x}
  \langle v_{1 x} v_{1 x}(\tau) \rangle = \frac{1}{||\Omega||} \sum_\omega
  \sum_{n=1}^\infty\frac{\rho^n e^{-\rho}}{n!} \int d{\mathbf v}^{(n)}
  v_{1 x} v_{1 x}^* \prod_{i=1}^n \phi({\mathbf v}_i) \; .
\end{equation}

The $x$ component of the post-collision velocity $v_{1 x}^*$ may
be written using Eq.~(\ref{eq-collision}) for the $\pm \pi/2$
collision rule discussed above as
\begin{eqnarray}
  v_{1 x}^* &=& \frac{1}{4\pi}\int d\hat{n} \Big( \mathsf{V}_x
  +\hat{n}_x\left[
  \hat{n}\cdot\left(\mathsf{V}^{(\alpha)}-\mathsf{V}\right)
   \right] \Big) \nonumber \\
   &&+\frac{1}{(4\pi)^2}\int d\hat{n}^{(\alpha)}\int d\hat{n}\;
   \hat{n}^{(\alpha)}_x  \nonumber \\
  &&\quad \times  (\hat{n}^{(\alpha)}\cdot \hat{n})\,\Big[\hat{n} \cdot
   \left(\mathbf{v}_1-\mathsf{V}^{(\alpha)}\right)\Big]
    \; ,
\end{eqnarray}
where, $\hat{n}$ and $\hat{n}^{(\alpha)}$  are the normal vectors
associated with the rotation operators $\hat{\omega}$ and
$\hat{\omega}^\alpha$, respectively. As a result of this
integration we obtain,
\begin{equation}\label{eq-v1x}
  v_{1 x}^* = \frac{1}{3}\left(v_{1 x} + 2 \mathsf{V}_x\right)\; .
\end{equation}
Assuming that particles of different species have the same mass,
and substituting Eq.~(\ref{eq-v1x}) into Eq.~(\ref{eq-v1xv1x}), we
find,
\begin{eqnarray}
  \langle v_{1 x} v_{1 x}(\tau) \rangle &=&
  \sum_{n=1}^\infty\frac{\rho^n e^{-\rho}}{n!} \int d{\mathbf v}_1
  (v_{1 x})^2 \left(\frac{n+2}{3n}\right) \phi({\mathbf v}_1)
  \nonumber \\
&=& \frac{\langle v_{1 x} v_{1 x}\rangle}{3}
    \sum_{n=1}^\infty\frac{\rho^n e^{-\rho}}{n!}
    \left(\frac{2}{n} + 1\right) \; .
\end{eqnarray}
For large enough $\rho$, we may approximate this expression by
\begin{equation}\label{eq-v1xv1x-approx}
  \langle v_{1 x} v_{1 x}(\tau) \rangle \approx
  \frac{\langle v_{1 x} v_{1 x}\rangle}{3}
  \sum_{n=1}^\infty\frac{\rho^n e^{-\rho}}{n!}
  \left(2 + n\right) \; ,
\end{equation}
which yields,
\begin{equation}
r_D = \frac{2\left(1-e^{-\rho}\right) + \rho}{3 \rho} \; .
\end{equation}
Substituting $r_D$ in Eq.~(\ref{eq-diff}) the expression for the
diffusion coefficient is
\begin{equation}
D=\frac{k_bT}{2m}\left(
  \frac{2\rho + 1 - e^{-\rho}}{\rho - 1 + e^{-\rho}}\right)\;.
\end{equation}
This analytic formula is compared with the simulation results in
Fig.~\ref{fig-DvsRho} where it is seen that it provides an
excellent approximation to the simulation results over all of the
physically interesting density range.

\section{Reactive Dynamics} \label{sec:reaction}

Next, we consider a reactive system with $M$ finite-sized
catalytic spherical particles (C), and a total of $N = N_A + N_B$
A and B particles which react with the C particles through the
reactions,
\begin{equation}
  A + C \mathop{\rightleftharpoons}_{k_r}^{k_f} B + C \;.
\end{equation}
The A and B particles undergo both non-reactive and reactive
collisions with C, and the multi-particle collisions described in
Sec.~\ref{sec:multiCol} among themselves. The macroscopic mass
action rate law may be written as,
\begin{equation}
\frac{d}{dt} \delta \bar{n}_A(t)= -(k_f+k_r)\delta
\bar{n}_A(t)\equiv -k \delta \bar{n}_A(t)\;,
\end{equation}
where $\delta \bar{n}_A(t)=\bar{n}_A(t)-\bar{n}_A^{eq}$ is the
deviation of mean number density of A particles from its
equilibrium value, and $k=k_f+k_r$ is the reciprocal of the
chemical relaxation time. We have incorporated the fixed number
density of the catalytic C particles into the rate constants.

The microscopic evolution equation for this system may be written
by simply augmenting the free streaming evolution operator in
Eq.~(\ref{eq:freest}) with a Liouville operator $\mathsf{L}$ that
describes the interactions of the A and B particles with the C
particles. If the interactions of A and B with C are through
continuous potentials, $\mathsf{L}$ takes the standard form,
$\mathsf{L}= {\bf F}\cdot \mbox{\boldmath $\nabla$}_{\bf P}$,
where ${\bf F}$ is the force between the A and B particles and C
and ${\bf P}$ is the vector of the the momenta of the particles.

For the purposes of calculation and illustration, we adopt a model
where the C particles are fixed in space and have radius $\sigma$.
The A and B particles either bounce back from the catalytic
spheres without changing their identity or react with probability
$p_R$. In this case the evolution equation for any dynamical
variable in the system is given by
\begin{equation}
\frac{d }{dt}a({\bf X}^{(N)},{\bf V}^{(N)},t)= (\mathsf{L}_0 \pm
\mathsf{L}_{\pm} + \mathsf{C}) a({\bf X}^{(N)},{\bf V}^{(N)},t)\;,
\label{eq:aeq}
\end{equation}
where the $\pm$ signs apply for $t>0$ and $t<0$, respectively. The
Liouville operators $\mathsf{L}_{\pm}$ describing the reactive and
non-reactive collisions with the catalytic particles are given by
\begin{eqnarray}
\mathsf{L}_{\pm}&=&\sum_\alpha \sum_{j=1}^M \sum_{i=1}^N |{\bf
v}_{i}\cdot \hat{{\bf r}}_{ij}|\theta(\mp {\bf v}_{i}\cdot
\hat{{\bf r}}_{ij}) \delta({r}_{ij}-\sigma) (\hat{b}_{ij}-1)
\Theta^{\alpha}_i \nonumber \\
&&+ p_R \sum_\alpha \sum_{j=1}^M \sum_{i=1}^N
|{\bf v}_{i}\cdot \hat{{\bf r}}_{ij}|\theta(\mp {\bf v}_{i}\cdot
\hat{{\bf r}}_{ij}) \delta({r}_{ij}-\sigma) \nonumber \\
&&\quad \times \hat{b}_{ij}\Big(\gamma \mathcal{P}^{\alpha
\alpha'}-1\Big) \Theta^{\alpha}_i \;.
\end{eqnarray}
Here $\hat{{\bf r}}_{ij}= ({\bf x}_i - {\bf x}_j)/r_{ij}$ is a
unit vector along the line of centers between particle $i$ and the
catalytic sphere $j$, $r_{ij}= |{\bf x}_i - {\bf x}_j|$ is the
magnitude of this vector and the operator $\hat{b}_{ij}$ converts
the velocity of particle $i$ to its post-collision value after
collision with the catalytic sphere $j$,
\begin{equation}
\hat{b}_{ij} ({\bf v}_1,{\bf v}_2,\dots, {\bf v}_i,\dots,{\bf
v}_N)=({\bf v}_1,{\bf v}_2,\dots,{\bf v}_i^*,\dots,{\bf v}_N)\;.
\end{equation}
For bounce-back dynamics we have ${\bf v}_i^*=-{\bf v}_i$. The
operator ${\mathcal P}^{\alpha \alpha'}$ acts on the species
labels to effect reactive collisions so that ${\mathcal P}^{\alpha
\alpha'} \Theta_i^\alpha =\Theta_i^{\alpha'}$ where $\alpha'=B$ if
$\alpha=A$ and vice versa. The factor $\gamma$ accounts for the
possibility that the forward and reverse reactions occur with
different probabilities leading to an equilibrium constant
$K_{eq}= \gamma^{-1}$ which is different from unity.

\subsection*{Rate law}

The chemical rate law for this system my be derived by taking the
dynamical variable $a$ to be the deviation of the number of
particles of species A from its average value, $\chi=N_A
-<N_A>=\delta N_A=- \delta N_B$, where
\begin{equation}
N_A=\sum_{i=1}^N \Theta_i^A\;.
\end{equation}
The angular brackets $<\cdots >$ signify an average over an
equilibrium ensemble where the numbers of A and B molecules
fluctuate but their sum is fixed, $N_A+N_B=N$ Starting with
Eq.~(\ref{eq:aeq}) for $t>0$ and using standard projection
operator methods \cite{mori} we may write a generalized Langevin
equation for $\chi(t)$ in the form,
\begin{eqnarray}
&&\frac{d }{dt} \chi(t)= f_{\chi}(t)-\frac{<(\mathsf{L}_{-} \chi)
\chi>}{<\chi \chi>} \chi(t) \nonumber \\
&& \quad - \int_0^t dt' \; \frac{<(\mathsf{L}_{-} \chi)
e^{{\mathcal Q} {\mathsf L}_{+} t'}{\mathcal Q} {\mathsf
L}_{+}\chi>}{<\chi \chi>} \chi(t-t') \;, \label{eq:gle}
\end{eqnarray}
where we have introduced the projection operator ${\mathcal P} a=
<a \chi><\chi \chi>^{-1} \chi$ and its complement ${\mathcal
Q}=1-{\mathcal P}$. The random force is
$f_{\chi}(t)=\exp[{{\mathcal Q} {\mathsf L}_{+} t}]{\mathcal Q}
{\mathsf L}_{+}\chi$.

Averaging this equation over a non-equilibrium ensemble where
$\chi$ does not fluctuate yields the generalized chemical rate
law,
\begin{eqnarray}
&&\frac{d }{dt} \overline{\delta n}_A(t)= -\frac{<(\mathsf{L}_{-}
\chi) \chi>}{<\chi \chi>} \overline{\delta n}_A(t) \nonumber \\
&& \quad - \int_0^t dt' \; \frac{<(\mathsf{L}_{-} \chi)
e^{{\mathcal Q} {\mathsf L}_{+} t'}{\mathcal Q} {\mathsf
L}_{+}\chi>}{<\chi \chi>} \overline{\delta n}_A(t-t') \;. \nonumber \\
\label{eq:grl}
\end{eqnarray}

The contribution
\begin{equation}
\frac{<(\mathsf{L}_{-} \chi) \chi>}{<\chi
\chi>}=k_{0f}(1+K_{eq}^{-1})\;,
\end{equation}
determines the initial rate arising from direct collisions of the
$A$ and $B$ particles with the catalytic spheres. For bounce-back
collision dynamics of the A and B species with the catalytic
sphere C, we have
\begin{eqnarray}
k_{0f}=p_R \sigma^2 \Big(\frac{8 \pi k_B T}{m}\Big)^{1/2} n_C\;,
\label{eq:k0f}
\end{eqnarray}
where $n_C$ is the constant number density of catalytic spheres.
The memory term accounts for all diffusion-influenced effects
arising from recollisions with the catalytic spheres.

\section{Results} \label{sec:sim}

\subsection{Simulation method}

The simulation of model is carried out in a cubic box with sides
$L_B$ and periodic boundary conditions. The centers of the spheres
of radius $\sigma$ are located in this box, taking care to
preserve periodic conditions on the edges when the spheres lie
partially outside the cube. Once the catalytic spheres are placed
in the box, the initial positions of the particles are assigned
values that are within the cube but outside the spheres. The
velocities are chosen from a Maxwell-Boltzmann distribution.

Given the initial distribution of particles and particle
velocities, the simulation begins by grouping the particles in
cubic cells of size $1$ within which the multi-particle collision
operators act to change the velocities of all particles,
preserving their positions. Then the displacement of each particle
is computed using the post-collision velocity, taking into account
the periodic boundary conditions of the cube and the bound-back
collisions with the spheres. When a particle hits a sphere it may
react with probability $p_R$, and the sign of its velocity is
changed. Collisions between particles and spheres occur in
continuous time in the interval $[ t , t+\tau ]$. When many
catalytic spheres are present a particle may hit several spheres
in one unit time $\tau$.

Once all the particles have been moved, the time advances one unit
$\tau$ and the particles are regrouped to apply the multi-particle
collision rule again.

\subsection{Single catalytic sphere}

In order to test the utility of the mesoscopic model we
investigate a system that contains a dilute distribution of
independent catalytic C particles so that the dynamics may be
described by considering a single C particle (labelled 1) with
radius $\sigma$ in a medium of A and B particles. In the case
where A particles are converted irreversibly to B upon collision
with C the chemical rate law takes the form, $d
\overline{n}_A(t)/dt =-k_f(t)\overline{n}_A(t)$, where $k_f(t)$ is
the time dependent rate coefficient. If the dynamics of the A
density field may be described by a diffusion equation, we have
the standard partially absorbing sink problem first considered by
Smoluchowski. \cite{smol} To determine the rate constant we must
solve the diffusion equation
\begin{equation}
\frac{\partial n_A({\bf r},t)}{\partial t}= D_A  n_A({\bf r},t)\;,
\end{equation}
subject to the boundary condition \cite{ck}
\begin{equation}
4 \pi D \bar{\sigma}^2 \hat{{\bf r}}\cdot (\mbox{\boldmath
$\nabla$}n_A)(\hat{{\bf r}}\bar{\sigma},t)=k_{0f} n_A(\hat{{\bf
r}}\sigma,t)\;.
\end{equation}
This equation assumes that the continuum diffusion equation is
valid up to $\bar{\sigma}> \sigma$, which accounts for the
presence of a boundary layer in the vicinity of the the sphere
surface where the continuum diffusion description should fail. The
resulting expression for the time-dependent rate coefficient is
\cite{advcp}
\begin{eqnarray}
k_f(t)&=&\frac{k_{0f} k_D}{k_{0f}+k_D} \nonumber  \\
 & +&\frac{k_{0f}^2}{k_{0f} +k_D}
 \exp\Big[\Big(1+\frac{k_{0f}}{k_D}\Big)^2
 \frac{D}{\bar{\sigma}^2}t\Big] \nonumber \\
 && \times {\rm erfc}\Big[\Big(1+\frac{k_{0f}}{k_D}\Big)
 \Big(\frac{Dt}{\bar{\sigma}^2}\Big)^{1/2}\Big]\;.
 \label{eq:dkoft}
\end{eqnarray}
Here $k_D=4 \pi \bar{\sigma} D$ is the rate constant for a
diffusion controlled reaction for a perfectly absorbing sphere.

\begin{figure}[htbp]
\centerline{\mbox{
\epsfig{file=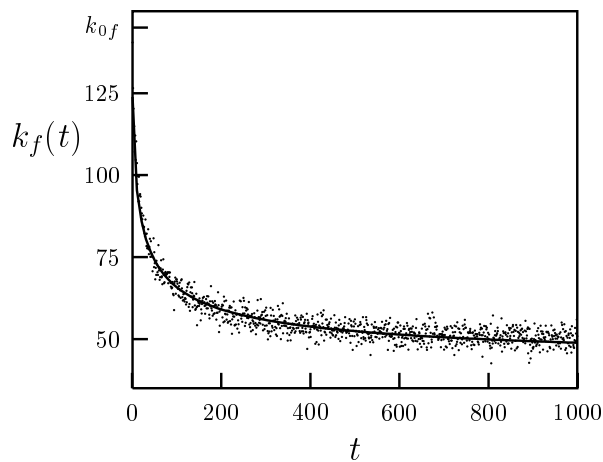,width=0.75\linewidth,clip=,angle=90}
}} \caption{Plot of the time dependent rate constant $k_f(t)/n_C$
versus $t$ for $\sigma=10$. The solid line is theoretical value of
$k_f(t)$ using Eq.~(\ref{eq:dkoft}) and $\bar{\sigma}=\sigma+1$.
         \label{fig:ab-s10-deltaNavst}}
\end{figure}

The time-dependent rate coefficient $k_f(t)$ may be determined
directly from the simulation by monitoring the A species density
field and computing $-(d \overline{n}_A(t)/dt)/\overline{n}_A(t)$.
The results of such a computation for irreversible reaction
($\gamma=0$) with probability $p_R = 0.5$ is shown in
Fig.~\ref{fig:ab-s10-deltaNavst}.
The system size is $100^3$ volume units and there is a sphere of
radius $\sigma=10$ located in the center of the system. The
simulation starts with $N(0) = N_A(0) = 10^7$ particles of species
A with unit mass uniformly distributed in the space. The initial
velocities are Maxwell distributed with $k_BT/m = 1/3$. The time
dependent rate coefficient starts at $k_{0f}$ and decays to its
asymptotic value $k_f$. In our mesoscopic model the continuum
theory cannot apply on the scale of one multi-particle collision
cell, so we have taken $\bar{\sigma} =\sigma+1$ to approximately
account for the microscopic boundary layer. One sees good
agreement between the simulation and diffusion theory results.

\begin{figure}[htbp]
\centerline{\mbox{
\epsfig{file=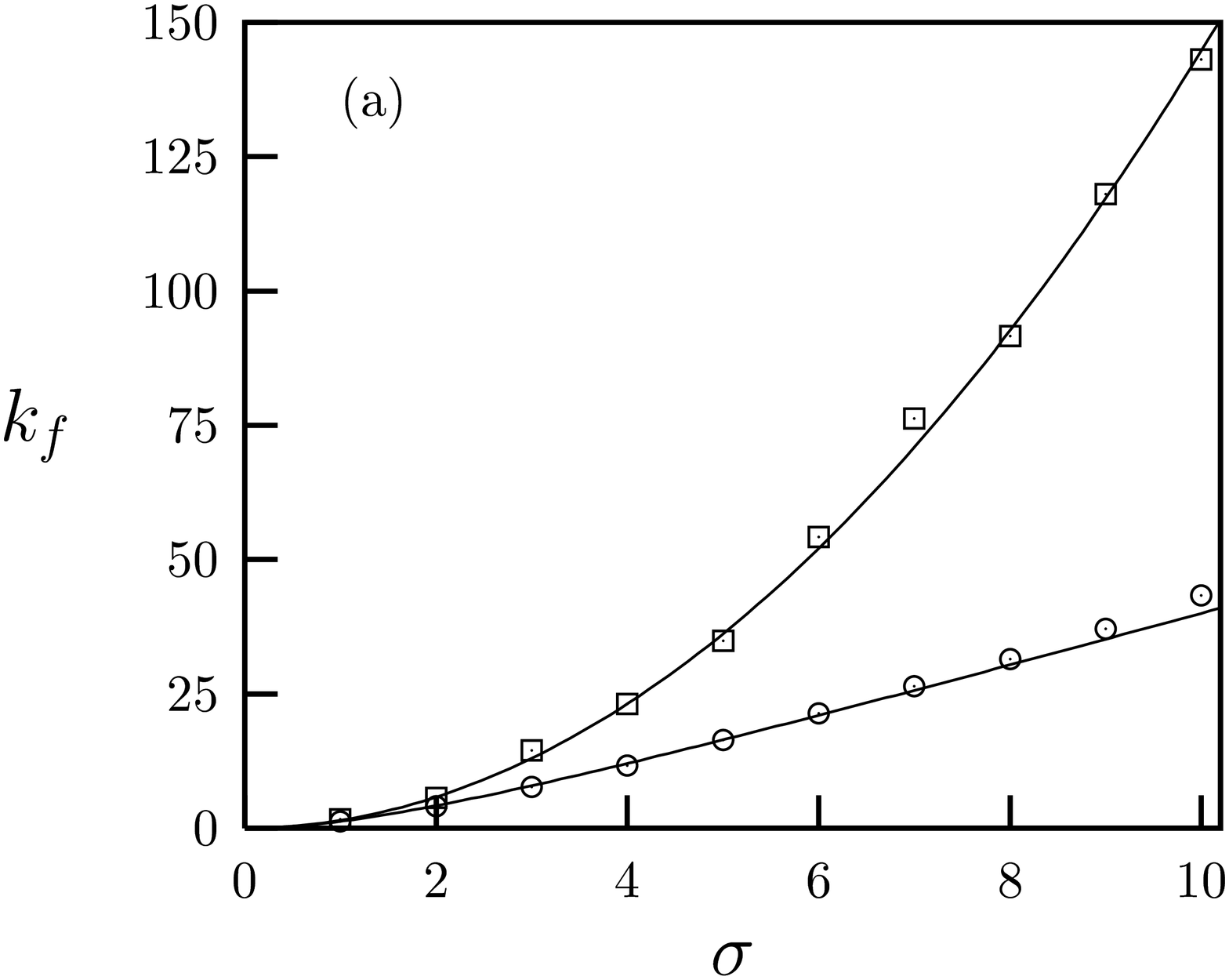,width=0.75\linewidth,clip=,angle=90}
}} \centerline{\mbox{
\epsfig{file=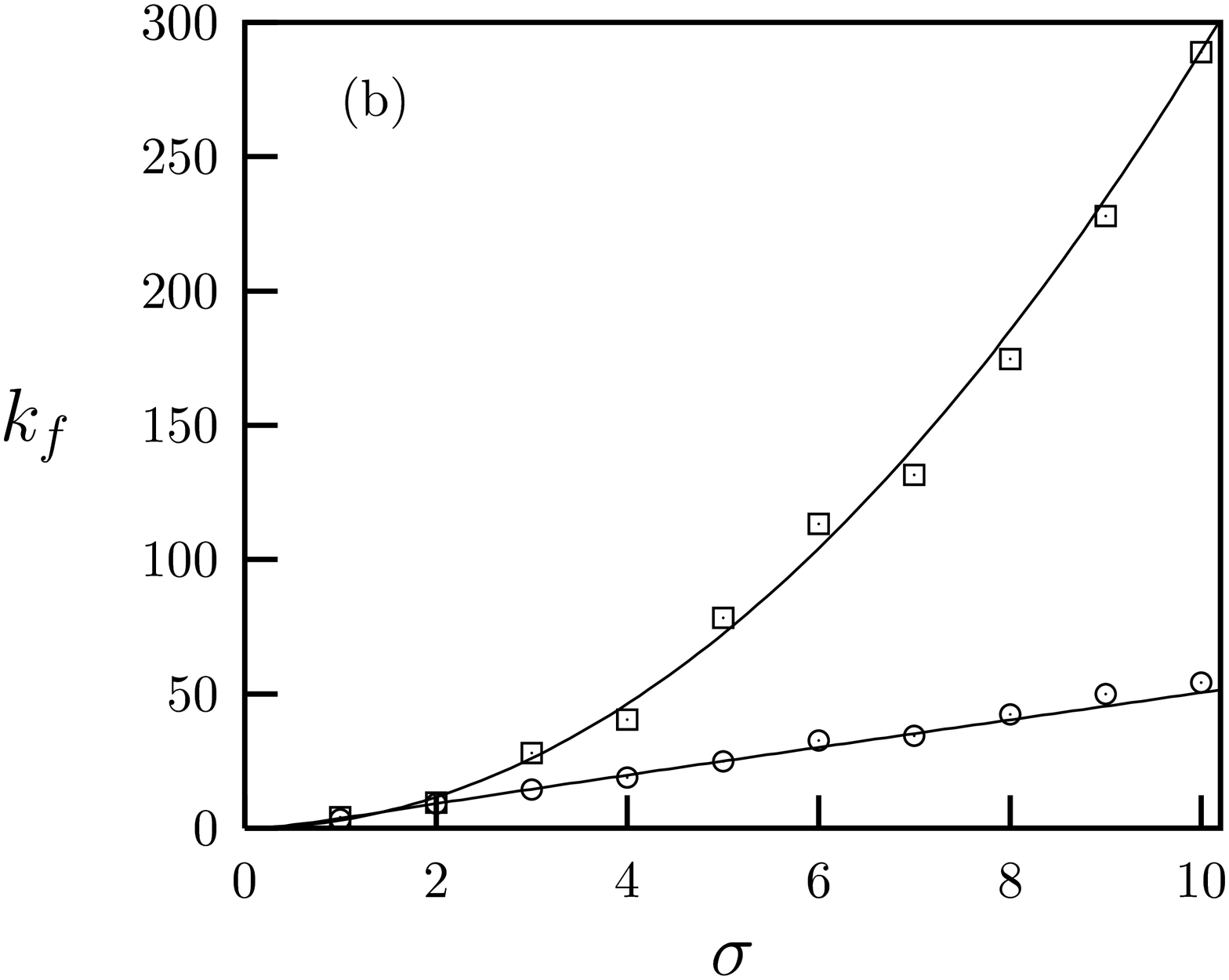,width=0.75\linewidth,clip=,angle=90}
}} \caption{Plot of $k_f/n_C$ ($\odot$) versus $\sigma$, the
radius of the catalytic sphere. The initial value
$k_f(t=0)=k_{0f}$ ($\boxdot$) is also plotted versus $\sigma$ in
this figure. The solid lines are the theoretical values of these
quantities determined from
$k^{-1}=(k_{0f}(1+K_{eq}^{-1}))^{-1}+k_D^{-1}$. (a) Irreversible
reaction ($K_{eq}^{-1}=0$) with $p_R = 0.5$. (b) Reversible
reaction ($K_{eq}^{-1}=1$) with $p_R = 1$.
         \label{fig:kfvsSigma}}
\end{figure}

In Fig.~\ref{fig:kfvsSigma}a we plot the values of $k_f$ extracted
from the simulation data in this way versus the radius of the
catalytic sphere.
The figure shows the increasing importance of diffusion-influenced
effects on the value of the rate constant as $\sigma$ increases.
While $k_{0f}$ grows quadratically with $\sigma$ in accord with
Eq.~(\ref{eq:k0f}), we see that $k_f$ grows more slowly and
approaches the diffusion-limited value of $k_D$, which depends
linearly on $\sigma$ for large $\sigma$. The theoretical estimate,
$k_f^{-1}=k_{0f}^{-1}+k_D^{-1}$, is in good agreement with the
simulation results.

A similar calculation can be carried out for the reversible case
($\gamma=1$ and $p_R=1$). For reversible reactions the chemical
relaxation rate $k(t)$ is given by Eq.~(\ref{eq:dkoft}) with
$k_{0f}$ replaced by $k_0=k_{0f}+k_{0r}=k_{0f}(1+K_{eq}^{-1})$
and, therefore, $k^{-1}=k_0^{-1}+k_D^{-1}$. \cite{pagitsas} For
our simulation conditions $K_{eq}=1$ so that $k_0=2k_{0f}$. Also
$k_f=k_r$. In Fig.~\ref{fig:kfvsSigma}b we plot the simulation
values of $k_f$ for the reversible reaction and compare them with
the diffusion equation formula. Once again good agreement is
found. The effects of diffusion appear at somewhat smaller values
of $\sigma$ since $k_0$ is larger for the reversible reaction and
the diffusion-limited value of the rate constant is reached at
smaller values of $\sigma$.

\subsection{Random distribution of catalytic spheres}

If instead of a single catalytic sphere we have a random
distribution of $M$ spheres of radius $\sigma$ in the volume $V$,
the rate constant will depend in a non-trivial way on the
catalytic sphere density or volume fraction $\phi=4 \pi \sigma^3
M/(3V)$. The reactions at one sphere surface will alter the A and
B particle density fields there. From the perspective of a
continuum diffusion equation approach, since the diffusion Green
function which couples the dynamics at the different spheres is
long ranged, the interactions from many catalytic spheres
determine the value of the rate constant. The problem is analogous
to the long range interactions that determine hydrodynamic effects
on the many-particle friction coefficient. There have been a
number of studies of the volume fraction dependence of the rate
constant
\cite{felderhof1,pagitsas,lebenhaft,felderhof2,felderhof3,gopich3,felderhof4}.
These derivations rely on resummations of classes of interactions
among the reacting spheres or other techniques.

The chemical relaxation rate for a system with a random
distribution of catalytic spheres with volume fraction $\phi$ is
given by
\cite{felderhof1,lebenhaft,pagitsas}
\begin{equation}
k(\phi)=k\Big[ 1+ \Big(
\frac{(k_{0f}+k_{0r})^3}{(k_{0f}+k_{0r}+k_D)^3}3\phi \Big)^{1/2} +
\cdots \Big]\;, \label{eq:conc}
\end{equation}
where, as earlier, $k^{-1}=(k_{0f}+k_{0r})^{-1}+k_D^{-1}$. The
first finite density correction to the rate constant depends on
the square root of the volume fraction. This non-analytic volume
fraction dependence arises from the fact that the diffusion Green
function acts like a screened Coulomb potential coupling the
diffusion fields around the catalytic spheres. As in the Debye
theory of electrolytes, one must sum an infinite series of
divergent terms to obtain the non-analytic $\phi$ dependence.

\begin{figure}[htbp]
\centerline{\mbox{
\epsfig{file=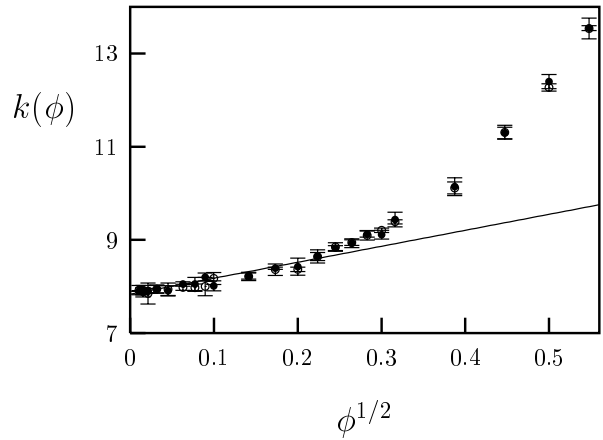,width=0.75\linewidth,clip=,angle=90} }}
\caption{Relaxation rate coefficient $k(\phi)/n_C$ as a function
of the square root of the volume fraction $\phi^{1/2}$ for
$\sigma=3$ and $k_BT=1/3$. Irreversible reaction $k_f(\phi)$
($\bullet$). Reversible reaction $k(\phi)$ ($\odot$).
For some values of $\phi$ the two cases cannot be distinguished in
the figure because the data points overlap. The solid line is
determined using Eq.~(\ref{eq:conc}).
         \label{fig:kvsPhi-S3}}
\end{figure}

The mesoscopic multi-particle collision dynamics follows the
motions of all of the reacting species and their interactions with
the catalytic spheres. Consequently, all many-sphere collective
effects are automatically incorporated in the dynamics. We have
carried out simulations of the chemical relaxation rate constant
$k(\phi)$ as a function of the volume fraction of the catalytic
spheres for a reversible reaction with $\gamma=1$
($K_{eq}^{-1}=1$) and $p_R=0.25$ as well as an irreversible
reaction with $\gamma=0$ ($K_{eq}^{-1}=0$) and $p_R=0.5$. For this
choice of parameters the theoretical formula predicts that
$k(\phi)$ for the reversible reaction is equal to $k_f(\phi)$ for
the irreversible reaction. Our simulations were performed for
systems with a volume fraction $\phi$ of catalytic spheres with
radius $\sigma=3$ in a system of size $100^3$ multi-particle cells
and an initial number density of $A$ particles, $n_A(0)=10$ per
cell. The results shown in Fig.~\ref{fig:kvsPhi-S3} were obtained
from an average over five realizations of the random distribution
of catalytic spheres.
We see that the simulation results confirm the existence of a
$\phi^{1/2}$ dependence on the volume fraction for small volume
fractions. As predicted by the theory for the chosen parameter
values the reversible and irreversible data overlap, even in the
high volume fraction regime. For larger volume fractions the
results deviate from the predictions of Eq.~(\ref{eq:conc}) and
the rate constant depends much more strongly on the volume
fraction. In this regime the diffusion coefficient is also
modified as a result of collisions with the catalytic spheres and
this effect also contributes to the deviation.

From these results we conclude that the mesoscopic multi-particle
collision dynamics provides a powerful tool for the exploration of
concentration effects on diffusion-influenced reaction kinetics.
Such concentration dependence is often difficult to explore by
other means.

\section{Conclusion} \label{sec:conc}

We have demonstrated that large-scale simulations of
diffusion-influenced reaction kinetics are possible by using the
mesoscopic multi-particle collision model. With this model the
dynamics of tens of millions of particles interacting with
hundreds of catalytic spheres could be followed for long times to
obtain the rate constants characterizing the population decay.
Such simulations would be very costly using full molecular
dynamics methods.

Since the dynamics is followed at the (mesoscopic) particle level,
a number of noteworthy features of the dynamical scheme are worth
mentioning. From a technical point of view the dynamics is stable
and no difficulties like those associated with discretizations of
the diffusion equation or boundary conditions arise. Reversible
and irreversible reaction kinetics may be treated in similar
fashion. All details of interactions arising from competition
among the catalytic spheres in a dense suspension are
automatically taken into account; thus, screening effects enter
naturally in the dynamics.

The model may be generalized to any reaction scheme and is not
restricted to the simple $A+C \rightleftharpoons B+C$ reaction
with catalytic C particles discussed in this paper. Since solute
molecules embedded in the mesoscopic solvent evolve by full
molecular dynamics (without solvent-solvent interactions), the
model will be most efficient when solvent-solvent interactions are
a major time limiting factor in the simulation. This could be case
for conformational changes of large molecules in solution,
reactions involving energy transfer in solution, etc. Thus, the
model should find applicability in a variety of circumstances when
diffusion-influenced reaction kinetics is important.

Acknowledgements: This work was supported in part by a grant from
the Natural Sciences and Engineering Research Council of Canada.

\end{document}